\documentclass[12pt,preprint]{aastex}
\usepackage{psfig,epsfig,amsfonts,amsmath,amssymb,graphicx,morefloats,appendix,tensor}
\usepackage{color}
\usepackage{natbib}
\usepackage{hyperref}
\begin{document}

\slugcomment{Published in ApJ: March 17, 2025}

\title{The Proxima Centauri Campaign -- \\ First Constraints On Millimeter Flare Rates from ALMA}

\author{Kiana Burton \altaffilmark{1}, Meredith A. MacGregor\altaffilmark{2}, Rachel A. Osten\altaffilmark{3,4}, Ward S. Howard\altaffilmark{1}, Alycia J. Weinberger\altaffilmark{5}, Evgenya Shkolnik\altaffilmark{6}, David J. Wilner\altaffilmark{7}, Jan Forbrich\altaffilmark{7,8}, Thomas Barclay\altaffilmark{9,10}}

\altaffiltext{1}{Department of Astrophysical and Planetary Sciences, University of Colorado, 2000 Colorado Avenue, Boulder, CO 80309, USA}
\altaffiltext{2}{Department of Physics and Astronomy, Johns Hopkins University, 3400 N Charles Street, Baltimore, MD 21218, USA}
\altaffiltext{3}{Space Telescope Science Institute, 3700 San Martin Drive, Baltimore MD 21218, USA}
\altaffiltext{4}{Center for Astrophysical Sciences, Department of Physics and Astronomy, Johns Hopkins University, 3400 North Charles St.,Baltimore MD 21218 USA}
\altaffiltext{5}{Earth \& Planets Laboratory, Carnegie Institution for Science, Washington, DC 20015, USA}
\altaffiltext{6}{School of Earth and Space Exploration, Arizona State University, Tempe, AZ 85287, USA}
\altaffiltext{7}{Center for Astrophysics $|$ Harvard \& Smithsonian, Cambridge, MA 02138, USA}
\altaffiltext{8}{Centre for Astrophysics Research, University of Hertfordshire, AL10 9AB, UK}
\altaffiltext{9}{NASA Goddard Space Flight Center, Greenbelt, MD 20771, USA}
\altaffiltext{10}{University of Maryland, Baltimore County, Baltimore, MD 21250, USA}

\begin{abstract}

Proxima Centauri (Cen) has been the subject of many flaring studies due to its proximity and potential to host habitable planets. The discovery of millimeter flares from this M dwarf with ALMA has opened a new window into the flaring process and the space-weather environments of exoplanets like Proxima b. Using a total of $\sim$~50 hours of ALMA observations of Proxima Cen at 1.3~mm (233~GHz), we add a new piece to the stellar flaring picture and report the first cumulative flare frequency distribution (FFD) at millimeter wavelengths of any M dwarf. We detect 463 flares ranging from energies 10$^{24}$~erg to 10$^{27}$~erg. The brightest and most energetic flare in our sample reached a flux density of 119 $\pm$ 7~mJy, increasing by a factor of $1000\times$ the quiescent flux, and reaching an energy of 10$^{27}$~erg in the ALMA bandpass, with t$_{1/2}$ $\approx$~16~s.  From a log-log linear regression fit to the FFD, we obtain a power law index of $\alpha_\mathrm{FFD}$ = 2.92 $\pm$ 0.02, much steeper than $\alpha_\mathrm{FFD}$ values ($\sim$2) observed at X-ray to optical wavelengths. If millimeter flare rates are predictive of flare rates at extreme-UV wavelengths, the contribution of small flares to the radiation environment of Proxima b may be much higher than expected based on the shallower power-law slopes observed at optical wavelengths. 

\end{abstract}

\section{Introduction}
\label{sec:intro}

\subsection{Stellar Flares and Habitability Concerns}

M dwarfs are considered the best chance to examine the habitability of exoplanets in the coming decade. Several factors make these stars ideal targets, including their prevalence as the most common spectral type in the galaxy \citep{Henry2006}, high occurrence rate of rocky planets in the habitable zone \citep{Dressing:2015, Shields2016, Hardegree:2019}, and relatively high signal-to-noise ratios of rocky planets compared to the host star. M dwarfs are cool, low mass stars with effective temperature ranges of $2400-4000$~K, and masses that span nearly an order of magnitude from $0.09 - 0.6$~M$_\odot$. The M dwarf spectral class crosses the fully convective boundary near M3 and $\sim$0.33~M$_\odot$ \citep{Chabrier:1997}, with different dynamo processes expected in the absence of a core for M4 and later dwarfs. These low-mass stars spin down more slowly than more massive stars \citep{Reiners:2012,Rebull:2018, Curtis:2019} and can remain relatively active up to 10~Gyr \citep{Curtis:2019,France:2020}. Because of these properties, M dwarfs have been known to produce considerably higher flare rates \citep{Davenport:2020,Davenport2016,Hawley:2014,Walkowicz:2011,Tristan:2023,Kowalski:2016} than stars of earlier spectral types. Photometric space-based missions like Kepler and the Transiting Exoplanet Survey Satellite (TESS) have enabled the extensive study of white-light flares from stars. \cite{Walkowicz:2011} found that stars that flare less frequently typically exhibit longer duration flares, while stars that have higher flare frequencies mostly release lower energy, short duration flares. They found that this trend was correlated to spectral type, with M dwarfs being the stars that flared most frequently at both higher and lower energies.  

Due to their high activity levels, many questions have been raised about M dwarfs' ability to host habitable planets \citep{Howard2018,Davenport2016,MacGregor:2021,Walkowicz:2011}. Stellar flares and their counterpart Coronal Mass Ejections (CMEs) and stellar energetic particles (SEPs) can be dangerous to nearby exoplanets, making the characterization of host star activity crucial for assessing habitability on nearby planets.  Frequent stellar flares can deliver damaging amounts of UV radiation that erode a planet's ozone layer by dissociating essential molecules like H$_2$O and O$_3$. Superflares (extreme events with estimated bolometric energies $>10^{33}$), are particularly dangerous for nearby planets with Earth-like atmospheres, as the ozone recovery rate is on the order of kiloyears \citep{Tilley2017}. However, \cite{Segura:2010} found that without considering the SEPs that accompany flares, atmospheric escape is not likely to occur, emphasizing the importance of quantifying the particle output of flares. 

Since different wavelengths probe different aspects of stellar flares, no one wavelength gives a complete picture of the physics at play. Flare Frequency Distribution (FFD) analyses have established that M dwarfs behave similarly at X-ray, near-ultraviolet (NUV), and far-ultraviolet (FUV) wavelengths (e.g., \citealt{Audard:2000, Robinson:2001, Mitra-Kraev:2005, Osten:2016, Berger:2024}). Dedicated multi-wavelength observing campaigns of EV Lac \citep{Osten:2005, Paudel:2021}, AD Leo \citep{Hawley:1991, Hawley:2003}, and AU Mic \citep{Tristan:2023} have detected multiple bright flares and enabled the extensive studies of the physics of stellar flares on M dwarfs.  However, multi-wavelength datasets are still extremely limited and have not included millimeter observations until recently (e.g., \citealt{MacGregor:2021, Howard:2022, Guns:2021}).  Although it is not possible to directly measure particle outputs across interstellar distances, we can probe the particles associated with flares through hard X-ray and radio emissions. Sensitivity limitations often prevent us from using hard X-ray observations, making observations of stellar flares at radio (microwave--centimeter) wavelengths extremely crucial for providing direct constraints on the energetic particle output of a star \citep{Beasley&Bastian:1998,Klein:2017}. The recent detections of millimeter flares (at 233~GHz) from Proxima Centauri (hereafter Proxima Cen), AU Mic, and $\epsilon$ Eridani \citep{Burton:2022}, have suggested that millimeter flaring emission might be a common aspect of stellar flaring missed until now.

\subsection{Proxima Centauri}

Proxima Cen is the closest exoplanetary system at a distance of 1.3020$\pm$0.0001~pc \citep{Gaia:2020} and hosts a terrestrial planet in its habitable zone at 0.0485~au (Proxima b) \citep{Anglada2016}. Proxima
Cen has also been well established as a highly active star (e.g., \citealt{Walker:2003, Davenport2016, Howard2018,Vida:2019, MacGregor:2021, Zic:2020}), making it a prime target to investigate the effects of stellar activity on the habitability of planets
orbiting M dwarfs. Low to moderate energy flares have been detected from Proxima Cen across multiple wavelengths. This includes events with energies up to 10$^{31.5}$~erg in the MOST bandpass \cite[4500-7500~Angstroms,][]{Davenport2016}, and 10$^{32}$~erg in the X-ray \citep{Gudel2004}. \cite{Howard2018} detected the first superflare with energy 10$^{33.5}$~erg in the Evryscope $g'$ bandpass, where the star increased $68\times$ its quiescent brightness. In addition, Evryscope has recorded other large flares with bolometric energies ranging from 10$^{30.6}$ to 10$^{32.4}$~erg \citep{Howard2018}. \cite{Davenport2016} estimate $\sim$~8 flares in the optical region occur each year with energy $\ge$10$^{33}$~erg. Highly circularized radio flares at 1.6~GHz have also been detected from Proxima \citep{Torrez:2021}. \citet{MacGregor:2018a} unexpectedly detected the first millimeter flare from Proxima Cen, after re-analyzing archival observations from the Atacama Large Millimeter/submillimeter Array (ALMA). 

The Proxima Cen Campaign \citep{MacGregor:2021} is the first multi-wavelength campaign to include simultaneous observations of Proxima Cen at millimeter wavelengths, giving us a new look at the physics behind stellar flares. This campaign resulted in observations that spanned from the radio to the X-ray, using Australian Square Kilometre Array Pathfinder (ASKAP), ALMA, Hubble Space Telescope (HST), TESS, the du Pont Telescope Las Cumbres Observatory Global Telescope (LCOGT), Swift, and Chandra, with the goal of providing more insights on the physics of stellar flares.  Here, we present the results from the ALMA data taken as part of this campaign. We have included all of the available ALMA observations ($\sim$50 hours) of Proxima Cen to measure the first FFD at millimeter wavelengths of any M dwarf. Section \ref{sec:observations} discusses the details of the observations from both the campaign and ALMA archive. In Section \ref{sec:analysis}, we explain our pipeline and analysis for characterizing time-variable emission from ALMA and obtaining observable flare properties at this wavelength. In Section \ref{sec:disc}, we explore the connection between flares at millimeter and optical wavelengths, interpret our results, and discuss their implications. In section \ref{sec:conclusion}, we summarize and discuss needed future work.

\section{Observations}
\label{sec:observations}

To create a complete ALMA dataset for Proxima Cen, we combined archival observations taken in 2017 and 2021 with observations from the Proxima Cen Campaign taken in 2019. Observations from 2017 and 2019 include four spectral windows with a total bandwidth of 2~GHz each (8~GHz total bandwidth) centered at 225, 227, 239, and 241~GHz. The correlator set-up maximized continuum sensitivity, and the XX and YY polarizations were obtained for all data (2017, 2019, and 2021). Additional details on each dataset are provided below. All analysis was performed using the Common Astronomy Software Package \cite[\texttt{CASA}, version 6.4.3.27, ][]{McMullin:2007}.  Imaging made use of the \texttt{tclean} task.

\subsection{2017 Observations}
ALMA observed Proxima Cen (PI: Anglada, 2016.A.00013.S) with both the full 12-m array and the Atacama Compact Array (ACA, comprised of 12 7-m antennas). The star was observed with 50 antennas out of the full 12-m array for two scheduling blocks (SBs) and 13 SBs with 8--11 antennas from the ACA. Each SB was split into `scans,' or 6.58 minute integrations alternating with the following phase calibrator observations: J1424-6807 or J1329-5608. The integration time for these observations was 1 second for the ACA observations and 2~s for the 12-m observations. Absolute flux (Ganymede, Callisto, Titan, J1427-4206, and J1517-2422) and bandpass (J1427-4206 and J1924-2914) calibration were performed. The precipitable water vapor (PWV) conditions were good ranging from $0.58-1.74$~mm. Additional discussion of these observations can be found in \cite{Anglada2016}.

\subsection{2019 Observations}

Proxima Cen was observed over 11 days between April and July of 2019 for a total of 34 hours of on-source time with the ACA as part of the Proxima Cen Campaign (P.I. MacGregor, 2018.1.00470.S). Each day of observation was split into three to four SBs that each lasted over an hour. Each SB was split into scans of 6-7 minutes that alternated with a phase calibrator. All data were taken with an integration time of 1 second. The quasars used for all calibrations across the dataset are: J1517-2422, J1308-6707, J1326-5256, J1337-6509, J1424-6807, J1924-2914, J1524-5903, J1337-1257, J1617-5848, J1829-5813, and J1303-5540. Overall, observing conditions were good with a PWV of 1.8~mm. More information about these observations including the number of antennas, on-source time, and root mean squared (rms) noise is detailed in Table \ref{tab:tab1}. Addition details and initial results from a subset of the data can be found in \citet{MacGregor:2021} and \citet{Howard:2022}.

 \begin{deluxetable}{l c c c}
\tablecolumns{4}
\tabcolsep0.1in\footnotesize
\tabletypesize{\small}
\tablewidth{0pt}
\tablecaption{2019 Proxima Cen Campaign Observation Details\label{tab:tab1}}
\tablehead{
\colhead{Date} &
\colhead{Antennas} &
\colhead{On-Source Time} &
\colhead{rms}  \\
\colhead{} &
\colhead{} & 
\colhead{(min)} &
\colhead{($\mu Jy$)} 
}
\startdata
\centering
April 28 & 10 & 148.3 & 101  \\
April 29 & 9 & 197.8 & 122 \\
May 1 & 9 & 197.8 & 125\\
May 2 & 9 & 197.8 & 98.4 \\
May 3 & 9 & 172.5 & 103\\
May 6 & 10 & 197.8 & 147 \\
May 12 & 11 & 197.8 & 77.8  \\
June 25 & 10 & 180.6 & 129\\
July 12-13  & 11 & 148.32 & 115 \\
July 14-16 & 11 & 148.32 & 86.9 \\
July 16-17 & 11 & 296.68 & 118 \\
\enddata
\end{deluxetable}

\subsection{2021 Observations}
ALMA took four additional observations of Proxima Cen with the 12m array between $24-25$ March 2021 (PI Anglada, 2019.A.00025.S). Each observation consisted of $\sim$25 scans, each lasting $\sim$2 minutes. Three spectral windows were used at 228, 244, and 246~GHz and were set-up for continuum emission detection, while the fourth spectral window included the line CO $\nu$ = 0, J $\rightarrow$ 2  at 230.538 GHz. For the purpose of our analysis, we removed all channels with CO emission (channels $957-964$, $966-975$, $979-984$, $1007-1011$, $1063-1082$, and 1871) using the \texttt{CASA} task \texttt{mstransform}. The integration time for these observations was 2~s.  Observing conditions were good with a PWV of 1.285~mm.

\section{Results and Analysis}
\label{sec:analysis}

We have developed a pipeline that executes \texttt{CASA} tasks in a \texttt{python} environment in order to automate the deconvolution and analysis process. Our pipeline is capable of creating light curves and identifying flares, characterizing any millimeter flaring emission, and producing FFDs. We describe how our pipeline produces images and performs flux density measurements of the ALMA Band 6 observations in the following sections.

\subsection{Creating Light Curves and Identifying Flares}
\label{sec:lightcurve}

Our pipeline uses the \text{CASA} task \texttt{uvmodelfit} to fit point source models to the ALMA visibilities in order to determine the flux densities and their associated uncertainties. Models are fit to each integration of each scan in a particular observation, and the best-fit flux densities are recorded from the fits. We then produce flux density light curves and uncertainties from the model fits for each observation. 

Flare identification in the flux density light curves is carried out using a sigma-cut threshold process. Very bright flaring events will be noticeable immediately by eye in the light curves, while smaller events will be harder to distinguish from the noise. To detect flares with our pipeline, we define a signal to noise threshold of $3\sigma$, and require that flares span more than one integration in duration. Our threshold is defined assuming the rms noise follows Gaussian probabilities.  We therefore flag all integrations with flux densities higher than the 3$\sigma$ threshold as flare candidates for further inspection, with $\sigma$ being the rms noise of our data. 

We calculate the rms in the image plane by drawing regions away from the source using the CASA \texttt{imview} tool, and then determining the average background flux variation within the regions using the CASA task \texttt{imstat}. We define a file that encodes the locations of the regions that the noise should be averaged over; this is typically done by drawing regions far enough away from the source within the full primary beam. To exclude any background sources like background galaxies, or stars, we use the region files provided by Chittidi et al. (in prep.) for the 2019 observations. While we would expect uniform sensitivity across each scan within a given day, we noticed some scans were considerably noisier than others.  This is likely correlated with weather, or due to the fact that flaring emission will contaminate the rms noise, since the image is constructed from sampled visibilities that contain information about the sky brightness distribution everywhere.

To account for this variability and more robustly constrain the significance of flares, our pipeline calculates a representative rms noise for each scan. Within each scan, the representative rms noise is determined using the average noise of the three integrations with the lowest flux densities. The pipeline creates images of these integrations using the \texttt{CASA} task \texttt{tclean}, and uses the region files to compute the noise for each scan.  This eliminates contamination from scans with higher-than-normal noise (i.e., scans that occurred towards the end of the night, or during an antenna issue) and from integrations with flaring emission. Using our list of representative rms values and significance threshold, our pipeline generates a list of flaring candidates, and discards all other events as spurious noise. From this list, the pipeline identifies all flaring candidates that are within the integration time as part of the same event to avoid the presence of duplicate flare events. The pipeline then records the flare peak, as well as the flare start and stop times. We note that the pipeline may not estimate the start and stop times of low energy flares efficiently. For example, if the true flare start and/or stop times fall below our pipeline's detection threshold, they will instead be recorded as the first and last integration that peaked above 3$\sigma$. We report a total of 463 flaring events. All large flares detected above 8$\sigma$ are listed in Table \ref{tab:tab2}.

\begin{figure}[t]
  \begin{center}
       \includegraphics[scale=0.55]{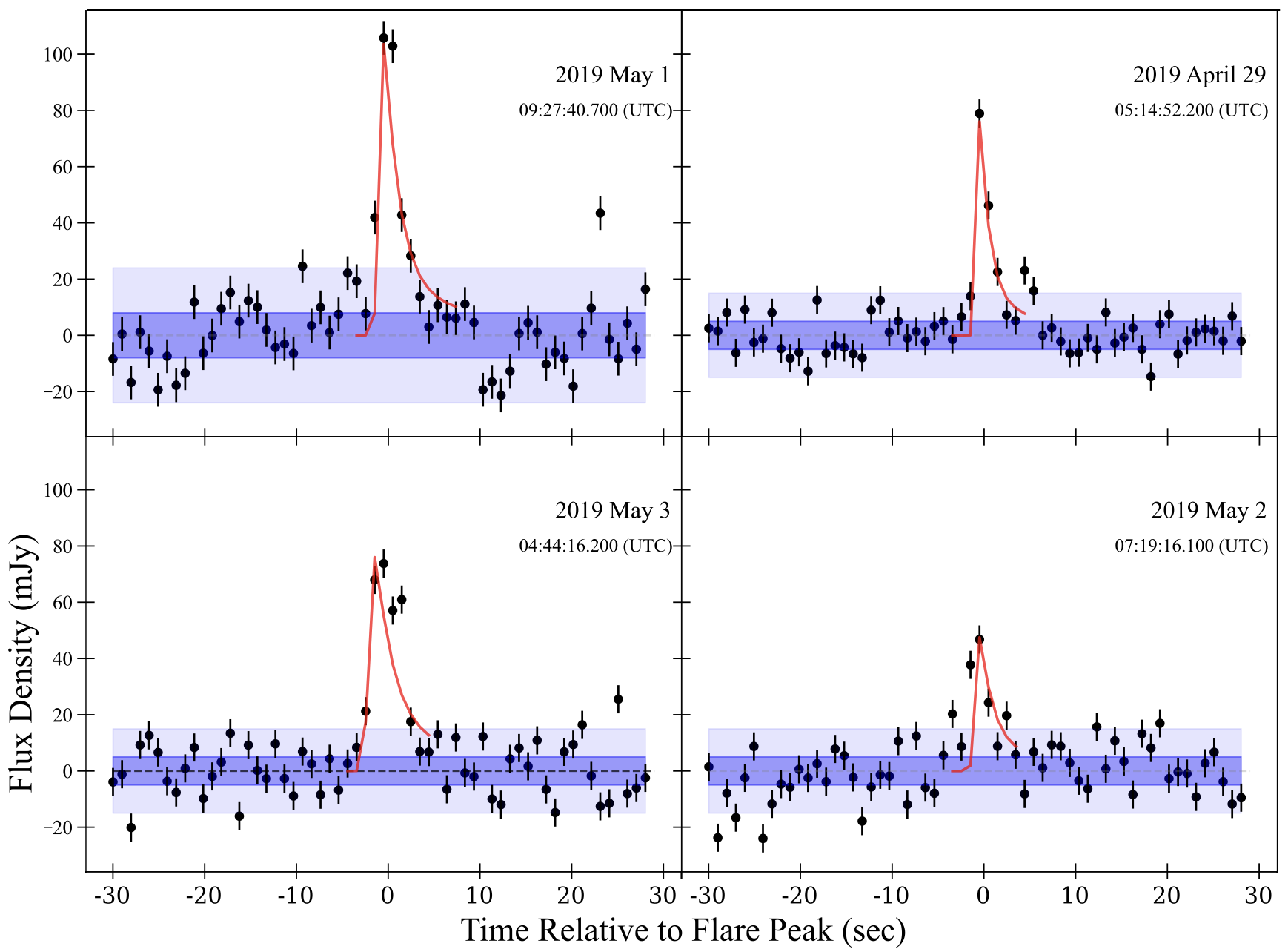}
  \end{center}
\caption{\small Panel of light curves produced for four energetic flares detected in our sample. The light curves are from flares detected from the Proxima Cen Campaign (2019 observations). All data are plotted with the same y-axis range for comparison. The dark and light blue shaded regions represent $\pm1\sigma$ and $\pm3\sigma$, respectively. The exponential fits to the flares (parameters and $\chi^2$ values are provided in Table~\ref{tab:tab2}) are indicated by the red lines.}
\label{fig:fig1}
\end{figure}

\begin{figure}[t]
  \begin{center}
       \includegraphics[scale=0.55]{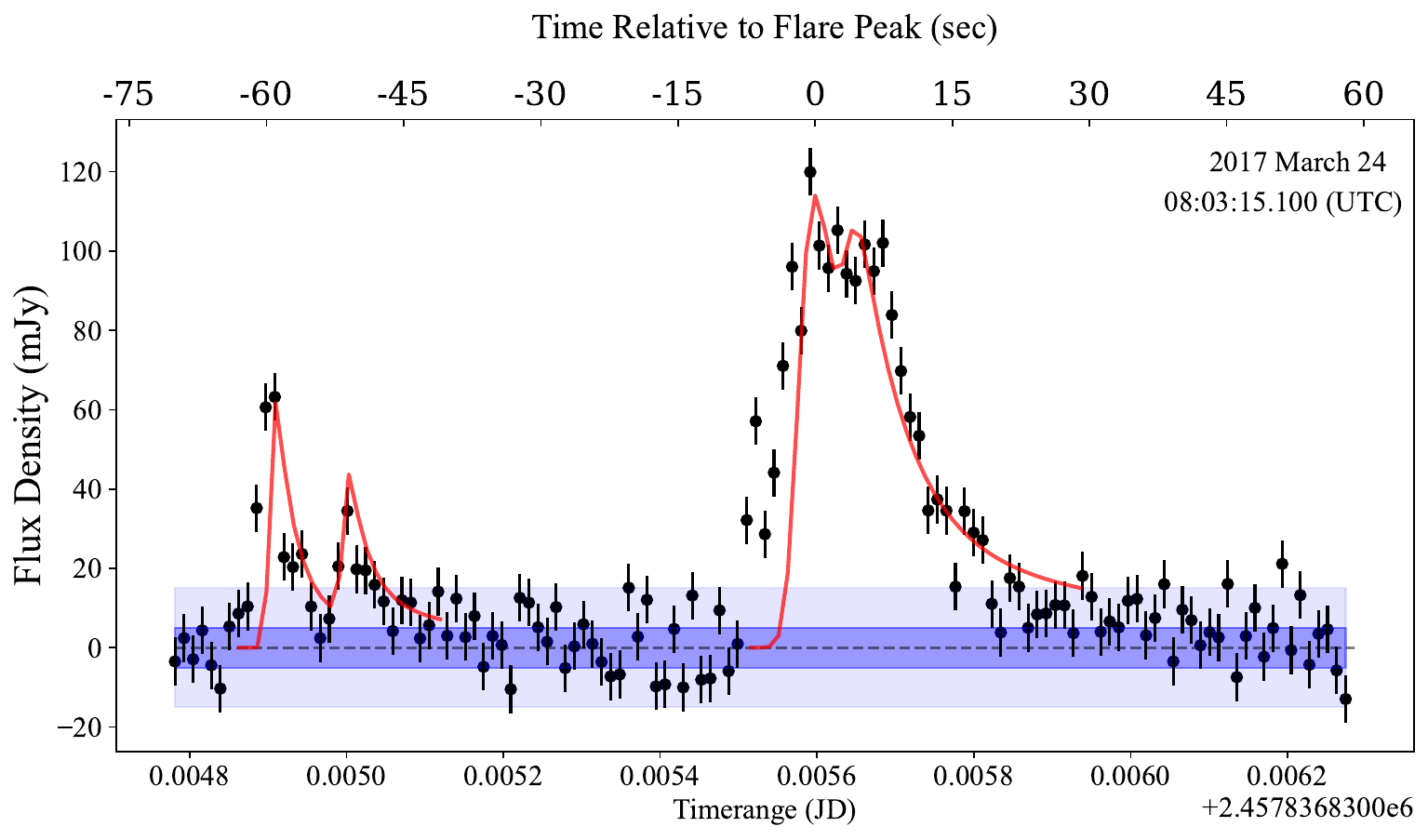}
  \end{center}
\caption{\small Light curve of the March 24th event, shown with two-component exponential fits. The dark and light blue shaded regions represent $\pm1\sigma$ and $\pm3\sigma$, respectively.} 
\label{fig:fig2}
\end{figure}

Figure~\ref{fig:fig1} shows examples of the flare light curves and typical noise backgrounds. We plot a gray dashed line at 0 for each light curve along with $\pm\sigma$ (dark blue) and $\pm3\sigma$ (light blue) shaded regions to illustrate how the data oscillate between the noise and detection thresholds of our pipeline. These are the most energetic and temporally resolved flares detected from our data and are well above 3$\sigma$. The largest flare detected from Proxima Cen is shown in  Figure~\ref{fig:fig2}, and occurred on 24 March 2017. This flare reached a peak flux density of $119\pm7$~mJy in the ALMA bandpass. A smaller flare that occurred $\sim$ 60~s before the onset of the larger flare is also seen in this light curve. We note that the 1 May 2019 and 24 March 2017 flares have already been published at a lower time resolution that did not fully resolve the flare peak structure present in our 1 s cadence light curves \citep{MacGregor:2021,MacGregor:2018a}, and thereby reported a lower peak flux density. The other flares shown in Figure~\ref{fig:fig1} have never been reported and are new results from the Proxima Cen Campaign executed in 2019. Their peak flux densities, luminosities, and energies are listed in Table~\ref{tab:tab2}.

\subsection{Flare Energy Calculations}
We calculate the flare energy in the ALMA bandpass using the following equation

\begin{displaymath}
E = 4 \pi d^2  \int_{t_{0}}^{t_{f}} \int_{v}^{}{F(t,\nu) dt d\nu}
\end{displaymath} 

\noindent where $d$ is the distance \cite[1.3~pc for Proxima Cen,][]{Lurie:2014},  $F(t,\nu)$ is the flux at each integration of the flare, $dt$ is the integration time in seconds, $t_0$ and $t_f$ are start and stop times of the flares, and  $\nu$ 
is the observing frequency bandwidth. This gives us the energy in ergs in the ALMA bandpass. The flares in our sample range in energy from 10$^{24}$~erg to 10$^{27}$~erg.  

To calculate the flare energy, we assume that the emission is isotropic, since we have no additional constraints on any directivity.  An anisotropic pitch-angle distribution could affect the emission intensity \cite[e.g.,][]{Fleishman:2003}, and motivates future modeling efforts. We also assume a constant flux density versus frequency within ALMA Band 6 because we are only able to weakly constrain the spectral indices for even high signal-to-noise events. We have explored the effect of this assumption on our flare energy calculations by using various spectral indices and have determined that this makes little difference from assuming a constant variation of flux density with frequency. Separating the flux dependence on time and frequency and considering the full range of spectral indices reported in Table~\ref{tab:tab2}, the variation in derived energy relative to the flat spectrum assumption is only $0.1\%-7\%$, which is smaller than the measurement errors on the individual flux points.

\subsection{Flare Frequency Distribution}

\begin{figure}[t]
  \begin{center}
       \includegraphics[scale=0.6]{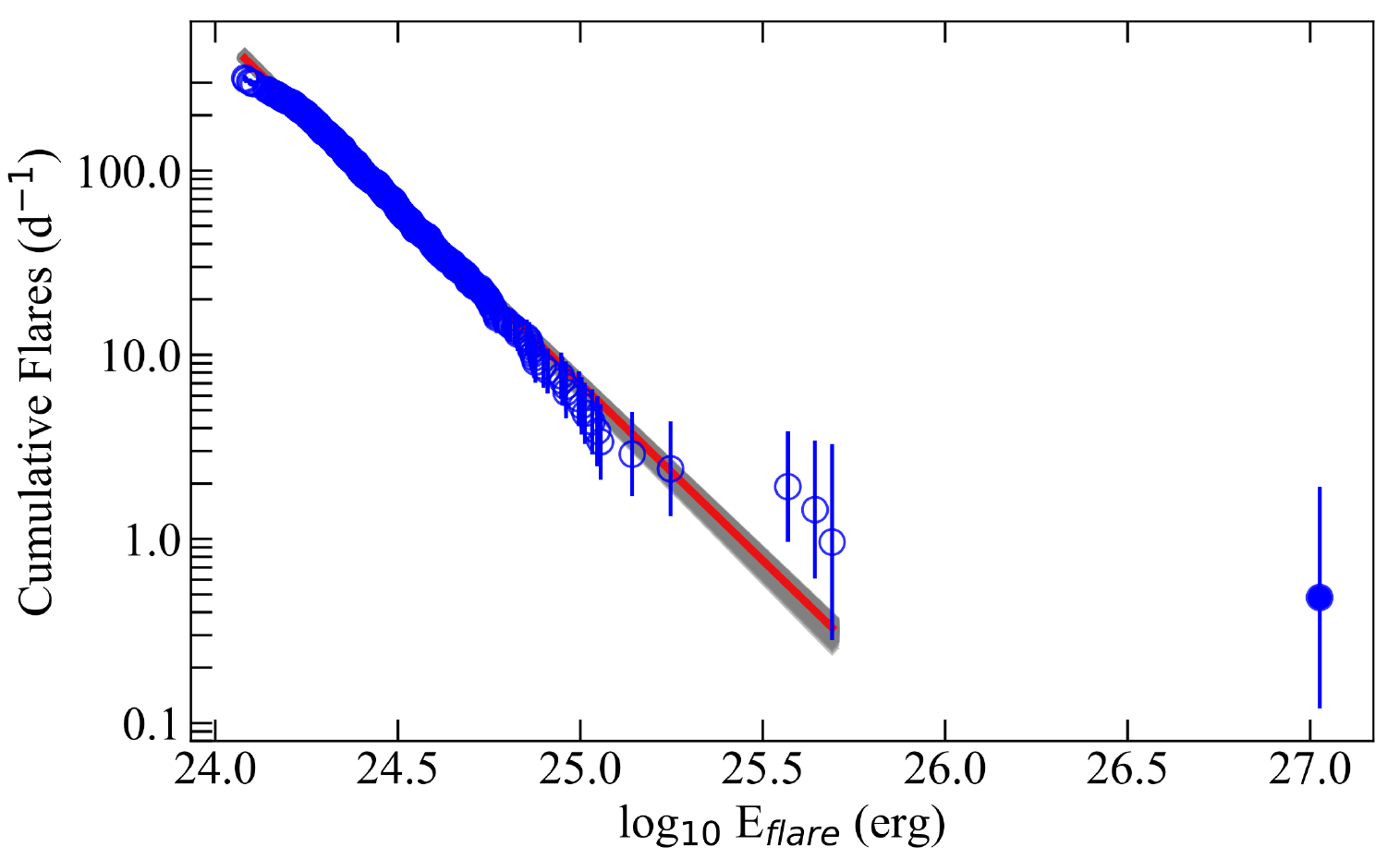}
  \end{center}
\caption{\small FFD of all of the available Proxima Cen ALMA Band 6 observations (463 flares). The energy reported is computed in the ALMA bandpass.  A log-log linear regression is performed to fit the FFD, yielding the parameters $\alpha_{cum}$ = -1.92 $\pm$ 0.02 and $\beta$ = 53 $\pm$ 0.31. The best-fit result is represented by the red line, with the gray shaded region indicating the uncertainty.}  The best-fit value is shown by the red line with the gray shaded region indicating the uncertainty. The errors in the cumulative occurrence rates calculated from 1$\sigma$ Poisson errors are plotted with the error bars. The largest flare is excluded from the fit and indicated by the shaded marker.
\label{fig:fig3}
\end{figure}

The FFD (shown in Figure~\ref{fig:fig3}) follows the probability distribution given by \citet{Lacy:1976}: 

\begin{displaymath}
dN(E) = kE^{-\alpha_\mathrm{FFD}}dE,
\end{displaymath} 

\noindent where $N$ is the number of flares that occur within the given time period for the flare energy $E$, $k$ is the constant of proportionality, and $\alpha_\mathrm{FFD}$ is the power law index.  Here, we only consider the flare energy in the ALMA bandpass and do not compute the total bolometric flare energy. Integrating the equation and expressing the relationship as a logarithm results in the standard cumulative FFD:

\begin{displaymath}
\text{log}_{10}(\nu) = \beta + \alpha_\mathrm{cum}\text{log}_{10}(E),
\end{displaymath} 

\noindent where $\nu$ is the cumulative occurrence for energies above some value, $\beta$ = $log_{10}\frac{k}{(1+\alpha_\mathrm{FFD})}$, and $\alpha_\mathrm{cum}$ = 1 - $\alpha_\mathrm{FFD}$. We measure the cumulative FFD parameters $\alpha_\mathrm{cum}$ and $\beta$ using least squares linear regression and calculate the uncertainty in the cumulative occurrence rates using a Poisson 1$\sigma$ confidence interval statistic \citep{Gehrels:1986, Davenport2016}. We weight the fit by the Poisson uncertainties to avoid biasing it towards the rarer, high energy events. For the largest flare, the flare waiting time and our total observation time are comparable which can introduce significant bias in the calculated rate.  As a result, we exclude this outlier (indicated by the shaded marker in Figure~\ref{fig:fig3}) the overall fit. In order to estimate the uncertainty in the power law fit, we use 10,000 Monte Carlo posterior draws to obtain: $\text{log}_{10}(\nu) = (53 \pm 0.31) + (-1.92 \pm 0.02) \text{log}_{10} (E)$.  This fit gives a power law index of $\alpha_\mathrm{FFD} = 2.92 \pm 0.02$. We note that this FFD includes flaring events detected in 2017, 2019, and 2021.  Unfortunately, there were not a sufficient number of events detected in each year to allow us to consider them separately.  Long term monitoring of Proxima Cen does support a 7-year stellar activity cycle \citep{Wargelin:2017}, which we do not account for here and could affect the fit.

\subsection{Characterizing Millimeter Flaring Emission}
\label{sec:characteristics}

We examine the properties of the flares by characterizing the temporal behavior of events well-resolved in time, calculating spectral indices, and estimating the lower limit on the fractional linear polarization. To perform this analysis, we define a new significance threshold ($8\sigma$) and the pipeline isolates the flaring candidates at or above this level. This ensures that we have enough signal to noise, since these calculations require splitting the data over smaller bandwidths which effectively raises the rms noise. 

We follow the method of \cite{Kowalski:2013} to describe the time evolution of the flares by measuring the full width at half maximum ($t_{1/2}$) of the flare profiles. We limit these calculations to flares with longer durations ($\ge$ 5~s) for the most robust constraints on the flare timescales, giving us ($t_{1/2}$) values for 5 flaring events listed in Table \ref{tab:tab2}.  We note a trend between event duration and energy. This is consistent with what is found in other studies, where the energy and duration of the flare are correlated -- higher energy flares usually last longer than lower energy flares \citep{Hawley:2014,Mendoza:2022}. With lower energy flares, data points in the rise or decay phase of the event are likely to fall below our 3$\sigma$ detection threshold, making them undetectable by our pipeline. This leads to some inefficiencies in our pipeline's ability to accurately distinguish between multiple and single integration events at lower energies.  

The spectral index of a given flare, $\alpha_\mathrm{spec}$, describes the dependence of the flux on the frequency (F$_\nu\propto\nu^{\alpha_\mathrm{spec}}$). To calculate $\alpha_\mathrm{spec}$, the pipeline uses the \texttt{CASA uvmodelfit} task again to fit point source models to the upper ($230 + 232$~GHz) and lower sidebands ($219 + 217$~GHz) of the four spectral windows independently. In the Rayleigh-Jeans limit, we would expect the spectral index $\alpha_\mathrm{spec}$ = 2 for quiescent stellar emission. The spectral indices for all flares above 8$\sigma$ are listed in Table \ref{tab:tab2}. 

To examine any polarization characteristics of the flares,which would help constrain emission mechanisms, we take the fraction ($|Q/I|$) of the Stokes parameters $Q = <{E_{x}}^2>-<{E_{y}}^2>$ and $I = <{E_{x}}^2>+<{E_{y}}^2>$, where $E_x$ and $E_y$ are the flux densities determined from fitting point source models to the XX and YY polarizations independently. The true linear polarization fraction is given by $p_{QU}^2 = (Q/I)^2 + (U/I)^2 $, but we are unable to constrain the Stokes parameter $U$ since full polarization was not available for these observations. Thus, our calculation only represents the lower limit to the linear polarization fraction. $|Q/I|$ values at the flare peaks for all events above 8$\sigma$ are listed in Table \ref{tab:tab2}. We detect linear polarization signals for all of the events.

 \begin{deluxetable}{l c c c c c c c c c}
\tablecolumns{7}
\tabcolsep0.1in\footnotesize
\tabletypesize{\small}
\tablewidth{0pt}
\tablecaption{Millimeter Properties of Flares $\ge$ 8$\sigma$ \label{tab:tab2}}
\tablehead{
\colhead{Date} &
\colhead{Peak Flux Density } &
\colhead{Peak $L_R$} & 
\colhead{log$_{10}$ Energy } &
\colhead{$\alpha_\text{spec}$} &
\colhead{$|Q/I|$} &
\colhead{t$_{1/2}$ }\\
\colhead{} &
\colhead{mJy} &
\colhead{$10^{13} $erg $s^{-1}$ Hz$^{-1}$} & 
\colhead{erg} &
\colhead{} &
\colhead{} &
\colhead{sec}
}

\startdata
\centering
			1 Feb 2017 &50 & 10 & 24.8 & -0.56 $\pm$ 2.9 & 0.005 $\pm$ 0.01 & $\ddag$ \\
		1 Feb 2017 & 58 & 12 & 24.9 & 0.44 $\pm$ 2.5 & 0.601 $\pm$ 0.01 & $\ddag$ \\
		24 March 2017 &	63 & 13 & 25.8 & 1.9 $\pm$ 2.2 & -0.96 $\pm$ 0.006 & 5.9 $\pm$ 2;$^\text{a}$\\
    &	 & & & &  & 6.1 $\pm$ 2.3$^\text{a}$\\
		24 March 2017& 119 & 24 & 27.2 & -3.3 $\pm$ 1.1 & 0.44 $\pm$ 0.01 &  16 $\pm$ 4$^\text{a}$\\
   &	 & & & &  & 14 $\pm$ 3$^\text{a}$\\
		19 March 2017 &	39 & 8 & 24.7 & -16.2 $\pm$ 7.1 & 0.83 $\pm$ 0.01 & $\ddag$\\
  29 April 2019 & 78 & 16 & 25.7 & 2.3 $\pm$ 3.1 & -0.22 $\pm$ 0.02 & 3.0 $\pm$ 0.9$^\text{a}$\\
  1 May 2019 &106 & 21 & 25.7 & -2.13 $\pm$ 1.4 & -0.19 $\pm$ 0.007 & 4.7 $\pm$ 1.2 $^\text{a}$\\
  2 May 2019 & 47 & 9 & 24.9 & 2.25 $\pm$ 3.5 & 0.31 $\pm$ 0.02 & 4.3 $\pm$ 1.4 $^\text{a}$\\
3 May 2019 & 74 & 15 & 25.7 & -9.9 $\pm$ 1.6 &  -0.39 $\pm$ 0.01 & 5.9 $\pm$ 1.6 $^\text{a}$\\	
\enddata
\vspace{0.2cm}
$^\text{a}$These flares were fit with the flare templates from \cite{Mendoza:2022}.
The $\chi^2$ values resulting from the exponential fits to the flares are 43, 2.3, 0.03, 0.1, 1.2,  and 2.19 for the flares in chronological order.  The March 24th flares both required two components in order to fit the light curves.  The t$_{1/2}$ values are reported for both components individually, but the $\chi^2$ value reported above includes both components.  Although some of the $\chi^2$ values are poor, we chose to fit the flares with the same template for consistency in determining the t$_{1/2}$ values. 

\end{deluxetable}

\section{Discussion}
\label{sec:disc}

\subsection{Flare Frequency Distribution}

This is the first time that an FFD has been measured at millimeter wavelengths for any M dwarf. Particles accelerated along magnetic field loops release microwave to millimeter emission. We should see a connection between both the timings and intensities of white light and millimeter flares assuming that the precipitating particles heat the photosphere and produce the continuum optical emission.  FFDs have been measured previously for Proxima Cen at other wavelengths.  \cite{Vida:2019} reported a power law index of 1.81 $\pm$ 0.03 in the TESS bandpass (600-1000 nm), while \cite{Davenport2016} reported $\alpha_\mathrm{FFD}$ =  1.68 $\pm$ 0.1 in the MOST bandpass and 2.22 $\pm$ 0.26 in the Evryscope $g'$ bandpass. A power law index of 0.87 $\pm$ 0.3 at FUV wavelengths has been reported from \cite{Loyd:2018}. The $\alpha_\mathrm{FFD}$ values for white light flares from the larger population of fully convective M dwarfs are generally consistent with 2 \citep{Hawley:2014,Davenport2016,Mendina:2020,Ilin:2019}.  

The power law index derived from the fit to our millimeter FFD for Proxima Cen, $\alpha_\mathrm{FFD}=2.92\pm0.02$, is in disagreement with all of the previously measured values. This could imply a disconnect between the population of particles responsible for millimeter and optical emission. A similar disconnect was described by \cite{MacGregor:2021} for the large May 1st flare included in our larger sample here.  This flare showed a time delay ($\sim$~1 minute) between the flare peak as seen at optical and millimeter wavelengths, as well as differences in the relative intensities (0.9$\%$ in TESS bandpass as opposed to $>1000\times$ in the millimeter). 

The source of heating in the solar corona is still debated.  Smaller `nano-flares' \citep{Parker:1988} could provide sufficient energy if they occur at higher rate.  On the other hand, magnetoacoustic and Alfv\'{e}n waves could carry upwards through the chromosphere and corona before dissipating as heat \citep{Schatzman:1949}.  If flares are responsible for coronal heating, previous work suggests that $\alpha_\mathrm{FFD}$ $\le$ 2 if larger flares dominate, while $\alpha_\mathrm{FFD}$ values $\ge$ 2 are associated with a high contribution of  small flares \citep{Hudson:1991,Hannah:2008}.  Recent observational studies searching for nano-flares reach energies of $10^{21}-10^{24}$~erg and largely support these same trends \citep{Kuhar:2018,Upendran:2022}.  Given this, our measured $\alpha_\mathrm{FFD}$ indicates that the high frequency of low energy flares may dominate the flare contribution to coronal heating for Proxima Cen.  It is clear that the highest energy events in our FFD are not well-fit by the power-law that describes the low energy events. This could be because higher energy events do not follow the expected power law distribution that low to moderate energies follow, or it could simply be that these events are statistically rare. In the latter case, artificial inflation of the measured flare rate can occur when the total observation time and flare waiting time are comparable and only the shorter end of the waiting time distribution is sampled.

Solar flares observed in the EUV by \cite{Krucker:1998} exhibit power law indices between $2.3 - 2.6$, indicating that micro-flares contribute to the solar corona heating. \cite{Yashiro:2006} examined the differences between FFDs with and without CMEs and found that flares without CMEs show power law indices $\ge$ 2, supporting the fact that micro-flares may contribute to coronal heating.  \cite{Cranmer:2013} adapted a model of coronal loop heating using numerical simulations of solar field-line tangling and MHD turbulence to reproduce the excess central emission observed by ALMA from the M dwarf AU Mic, another well-known flare star \citep{MacGregor:2013}. If a steep power law index at millimeter wavelengths is observed for other M dwarfs, active coronal emission may be common.

\subsection{Temporal Behavior}
 All of the flares included in our sample have relatively short durations compared to flares observed at other wavelengths, with the longest $t_{1/2}$ values ranging from 3--16~sec.  Solar flares in the hard X-ray have been shown to undergo similar short bursts, with timescales ranging from 10--20~s \citep{Qiu:2012}. Other examples with short timescales include Type III bursts seen in LOFAR observations \cite[e.g.,][]{Morosan:2014}, NUV stellar flares \cite[e.g.,][]{Kowalski:2016}, and periodic variations during white-light flares on active, fully convective stars \cite[e.g.,][]{Mathioudakis:2006}.  In general, higher energy events appear to have longer decay phases than rises.  Two out of five of the events included in our analysis were already temporally characterized in \cite{MacGregor:2018a,MacGregor:2021}. The 2017 March 24th event was originally fit with a Gaussian and found to be mostly symmetric, with no pronounced exponential decay. Here, we show the same event at higher temporal resolution using 1~s integration times. With this higher resolution, the decay time does appear to be slower than the rise time.  To account for this, we fit all of the flaring events shown in Figures \ref{fig:fig1} and \ref{fig:fig2} with the empirical flare template with an exponential decay phase from \cite{Mendoza:2022}. Due to the complex temporal structure in the March 24th event, we fit this flare with a two-component model. We also fit the smaller flare which peaks $\sim$60~s before the main flare peak with a two-component model as is shown in Figure \ref{fig:fig2}. Most of the 2019 events are well fit by single exponential components. 

During the particle acceleration process, some particles will become trapped in the magnetic field loops later precipitating from the trap, while some will directly precipitate into the chromosphere. Radio (microwave to millimeter) emission is produced due to both trapping and precipitation. \cite{Lee:2002} separate the transport and acceleration effects of high energy particles during solar flares by dividing the particles into these two populations to interpret microwave and hard X-ray bursts. Their results show that flares with symmetric structure and little exponential decay are consistent with efficient precipitation. Similarly, \cite{MacGregor:2020} attributed the short timescales of millimeter flares to efficient particle precipitation and ineffective trapping from simple loop structures. In contrast, the majority of the millimeter flares we show here have longer decay phases and are well fit by exponential flare templates \citep{Mendoza:2022}. While the timescales of these flares are far too short to draw any meaningful insights from their temporal structures, the presence of exponential decay could indicate efficient trapping contrary to \cite{MacGregor:2020}. One flare in our sample (March 24th) has a significantly longer duration with a t$_{1/2}$ value of 16 $\pm$ 4~s. Although the overall fit quality for this event is poor, the complex structure of the flare could be attributed to a much more complicated loop--like structure.

\subsection{Polarization and Frequency Behavior}
By examining the spectral and polarization behavior of flares, we can probe potential emission mechanisms. Blackbody radiation in the Rayleigh-Jeans limit has a spectral index of $\sim$~2, so deviation from this value indicates other sources such as synchrotron or gyrosynchroton. The favored emission mechanism at microwave frequencies has been gyrosynchroton emission. Thus far, most millimeter flares detected from M dwarfs at 233~GHz have shown strictly negative spectral indices, implying optically thin emission \citep{MacGregor:2020}.  Due to limited sensitivity, we are only able to determine the evolution of spectral indices during the longer duration, high energy flares. For the May 1st and March 24th events, the evolution of the spectral indices and polarizations throughout the flares has been published previously in \cite{MacGregor:2021}.  Both events exhibit negative spectral indices.  For all other events, we are only able to constrain the spectral indices at the flare peaks. Some of these values are negative, but most have large uncertainties, making it hard to distinguish between negative and positive spectral indices. 

Intriguingly, positive spectral indices are observed for millimeter flares from the Sun \citep{Krucker:2013} and other Sun-like stars including $\epsilon$ Eridani \citep{Burton:2022}. In the Sun, these values range from 0.3 -- 0.5. In $\epsilon$ Eridani, three flares were detected at 233~GHz, with two out of three spectral indices being positive during the flares albeit with large uncertainties. Combined with our new results, this could indicate a difference between the emission mechanism for millimeter flares from different spectral type stars.  However, our spectral index calculations are low significance and make a definitive conclusion challenging. To better constrain the spectral indices and fully characterize the spectral behavior of these flares, we would need broadband coverage at microwave through millimeter wavelengths ($10-300$~GHz) with simultaneous observations from both the VLA and ALMA. Fitting both synchrotron and gyrosynchrotron SEDs to detected flares would test whether or not these events are truly tracing the optically thin part of the spectrum.

We detect linear polarization for all of the flares in our sample detected above 8$\sigma$. \cite{MacGregor:2020} used the polarization properties of the March 24th Proxima flare to distinguish between gyrosynchrotron and synchrotron emission mechanisms. They determined that synchrotron emission is most likely because of the linear polarization signals, but were unable to conclusively distinguish between the two emission mechanisms due to the prompt timescales of these flares. Synchrotron emission operates at higher harmonics of the electron gyrofrequency implying lower magnetic field strengths. However, fast bursts require higher particle densities, which implies strong magnetic field regions.  Unfortunately, all of the flares in our sample have similarly short timescales so we are still unable to distinguish between gyrosynchrotron and synchrotron emission mechanisms from these flares. The potential mixture of both positive and negative spectral indices also means that we are unable to pinpoint whether or not we are probing the optically thin or thick side of the gyrosynchrotron or synchrotron spectrum.

\section{Conclusions}
\label{sec:conclusion}

We have measured the first FFD at millimeter wavelengths for Proxima Cen. Our results indicate that flaring emission at millimeter wavelengths is common and emphasize the utility of using microwave to millimeter observations to gain new insights into stellar flares. Below we outline the takeaways from our analyses of the millimeter FFD and other observable properties of these flares:

\begin{enumerate}

    \item The measured millimeter FFD has a power law index $\alpha_\mathrm{FFD}$ = 2.92 $\pm$ 0.02, significantly steeper than what has been reported for most M dwarf FFDs at other wavelengths. This $\alpha_\mathrm{FFD}$ value indicates that low energy flares could dominate the coronal heating of Proxima Centauri.  Similar power law indices have been derived for observations of small solar flares, and support coronal heating due to nano-flares although wave heating is not excluded.  Our results are are also consistent with the \cite{Cranmer:2013} model of an active corona on AU Mic.
    
    \item Proxima Cen has been observed frequently at optical wavelengths, with a much shallower FFD power law index of 1.88 $\pm$ 0.06. This significant difference could indicate a disconnect between sources of optical and millimeter emission during flares. Since optical observations of stellar flares are more readily available and often used to infer the flaring flux at other wavelengths, this result underlines the need for further multi-wavelength campaigns to constrain scaling relations. In particular, the higher rate of millimeter flares compared to optical flares and the tight correlation between FUV and millimeter emission observed by \cite{MacGregor:2021} may suggest that the extreme-UV radiation environment of Proxima b due to small flares is also higher than predicted from the optical flare rate.
    
    \item  Unlike previous work that fit millimeter flares with a symmetric rise and decay phase, the majority of the events presented here are better fit with a standard stellar flare template that includes a longer decay phase.  This could be evident of efficient particle trapping.  Higher temporal resolution in future observations could further explore the exponential tail of millimeter flares. 
    
    \item The majority of our sample show evidence for negative spectral indices and linear polarization, which suggest that we are observing the optically thin part of either the synchrotron or gyrosynchrotron spectrum.  However, we are unable to conclusively distinguish between these two emission mechanisms. Multi-wavelength observations spanning a larger range of the microwave and millimeter spectrum (from 10 GHz to 230 GHz) are needed solve this puzzle. 
    
\end{enumerate}

Although this is the first complete analysis of a statistically significant population of millimeter flares, there is still much work to be done in order to fully explore the properties of flares at these wavelengths.  The simultaneous multi-wavelength data taken as part of the larger Proxima Cen Campaign will allow us to examine how millimeter emission correlates with optical through X-ray wavelengths \cite[e.g.,][]{MacGregor:2021,Howard:2022} in order to create a more complete picture of stellar flaring.  Proxima Cen remains an outlier given its high activity level for such an old star \citep{newton:2018}.  Going forward, similar studies must be executed for other M dwarfs with differing ages and spectral types to fully explore the expected stellar impact on planetary habitability. To achieve this, we are currently carrying out a broader millimeter flare monitoring campaign of $\geq$6 M stars of various ages and activity levels with ALMA (GO programs 2021.1.01209.S, PI: MacGregor and 2022.1.01163.S, PI: Howard), with simultaneous observations covering soft X-ray, near-UV, and optical wavelengths.

\section{Acknowledgements}

\vspace{1cm}
 This paper makes use of the following ALMA data: ADS/JAO.ALMA 2016.A.00013.S, 2018.1.00470.S, and 2019.A.00025.S. ALMA is a partnership of ESO (representing its member states), NSF (USA) and NINS (Japan), together with NRC (Canada) and NSC and ASIAA (Taiwan) and KASI (Republic of Korea), in cooperation with the Republic of Chile. The Joint ALMA Observatory is operated by ESO, AUI/NRAO and NAOJ. The National Radio Astronomy Observatory is a facility of the National Science Foundation operated under cooperative agreement by Associated Universities, Inc.   

 M.A.M. acknowledges support for part of this research from the National Aeronautics and Space Administration (NASA) under award number 19-ICAR19\_2-0041.  T.B. acknowledges support from the GSFC Sellers Exoplanet Environments Collaboration (SEEC), which is funded in part by the NASA Planetary Science Divisions Internal Scientist Funding Model.

\software{\texttt{CASA} \cite[6.4.3.27,][]{McMullin:2007}, \texttt{astropy} \citep{astropy:2013,astropy:2018}}

\bibliography{References.bib}

\end{document}